\documentstyle[multicol,aps,epsf]{revtex}
\newcommand{\beq}{\begin{equation}}
\newcommand{\eeq}{\end{equation}}
\begin{document}

\title{Fermion-Condensation Quantum Phase Transition
in High-$T_c$ Superconductors}

\author{M.Ya. Amusia$^{a,b}$ and V.R. Shaginyan$^{a,c}$
\footnote{E--mail: vrshag@thd.pnpi.spb.ru}}
\address{$^{a\,}$The
Racah Institute of Physics, the Hebrew University, Jerusalem 91904,
Israel;\\ $^{b\,}$Physical-Technical Institute, Russian Academy of
Sciences, 194021 St. Petersburg, Russia;\\ $^{c\,}$ Petersburg Nuclear
Physics Institute, Russian Academy of Sciences, Gatchina, 188350,
Russia}
\maketitle

\begin{abstract}
The effect of a quantum phase transition associated with the
appearance of the fermion condensate in an electron liquid on the
properties of superconductors is considered. It is shown that the
electron system in both superconducting and normal states exhibits
characteristic features of a quantum protectorate after the point of
this fermion-condensation quantum phase transition. The
single-particle spectrum of a superconductor can be represented by
two straight lines corresponding to two effective masses $M^*_{FC}$
and $M^*_{L}$.  The $M^*_{FC}$ mass characterizes the spectrum up to
the binding energy $E_0$, which is of the order of the
superconducting gap in magnitude, and $M^*_{L}$ determines the
spectrum at higher binding energies.  Both effective masses are
retained in the normal state; however, $E_0=4T$. These results are
used to explain the lineshape of single-particle excitations and
other remarkable properties of high-$T_c$ superconductors and are in
a good agreement with recent experimental data.  \end{abstract}


\begin{multicols}{2}

Recent experiments using angle-resolved photoemission electron
spectroscopy gave rather accurate data on the dispersion of
single-particle excitations over a wide range of binding energies
\cite{vall,blk,krc}. These experiments were carried out with high-$T_c$
superconductors Bi$_2$Sr$_2$CaCu$_2$O$_{8+\delta}$ differing in the
doping level both at temperatures $T$ below the critical temperature
$T_c$ of the decay of superconducting state and at $T_c\leq T$. It
was inferred that the dispersion of quasiparticle excitations
$\varepsilon({\bf p})$, where ${\bf p}$ is momentum, can be described
in the energy range (-200---0) meV by two straight lines intersecting
at the binding energy $E_0\simeq(50-70)$ meV \cite{blk,krc}. This
circumstance directly points to the existence of a new energy scale
in the self-energy part of quasiparticle excitations at temperatures
$T\leq T_c$ and $T_c\leq T$ \cite{blk}. Therefore, new additional
constraints can be imposed on the theories that are in principle
applicable to the description of properties of high-$T_c$
superconductors. For example, such scale is absent in the theories of
normal \cite{lan} and marginal \cite{var} Fermi liquids, as well as
in the theory based on an idea of the quasiparticle spin--charge
separation \cite{pwa}. The kink in the quasiparticle dispersion law
described above could be explained by the interaction of
quasiparticles with collective magnetic excitations
\cite{krc,norm} that was observed in high-$T_c$ superconductors in
experiments on inelastic neutron scattering at $T\leq T_c$ and
described in the literature; see, for example, \cite{fong}. However,
a dispersion kink is also observed at $T_c\leq T$, when these
collective excitations disappear. Moreover, these excitations are
successfully described as inelastic neutron scattering from Cooper
pairs \cite{abrn}, which is confirmed by other experimental results
\cite{he,abr}. With regard to this explanation of the physics of magnetic
excitations, it is unlikely that these excitations can
significantly affect the single-electron dispersion. Experimental
data on single-particle electron spectra of high-$T_c$ superconductors
with d-wave symmetry indicate that the perturbation of the
superconducting phase and single-particle spectra by phonons,
collective states, or impurities is very small. Therefore, this
state can be described as a strongly collectivized quantum state or
as ``quantum protectorate'' \cite{pwa,rlp,pa,msh}.
From here, it may be inferred that the kink description proposed in
\cite{krc,norm} is very likely contradictory to the quantum
protectorate concept.

In this paper, we show that the dispersion
kink can be explained based on the assumption that the electron
system of high-$T_c$ superconductor is after the point of the
Fermi-condensate quantum phase transition. Thus, the
fermion-condensation quantum phase transition (FCQPT) serves as the
point separating a normal Fermi liquid from a strongly correlated
liquid of a new type \cite{ks,vol} that fulfills the quantum
protectorate requirements.

Let us start with a brief description of the
properties of an electron system with the fermion condensate.
Consider a two-dimensional electron liquid in the
superconducting state at $T=0$ on a simple square crystal lattice,
which we replace temporarily by a uniform positive charge. Then, the
ground state energy $E_{gs}[\kappa({\bf p}),n({\bf p})]$
is a functional of the order parameter of
the superconducting state $\kappa({\bf p})$ and occupation numbers
$n({\bf p})$ \cite{vsl} and is determined by the known equation
\beq E_{gs}[\kappa({\bf p}),n({\bf p})]=E[n({\bf p})]\eeq
$$+
\int V_{pp}({\bf p}_1,{\bf p}_2)\kappa({\bf p}_1)
\kappa^*({\bf p}_2)
\frac{d{\bf p}_1d{\bf p}_2}{(2\pi)^4}.$$
The pairing interaction $V_{pp}({\bf p}_1,{\bf p}_2)$ is assumed to be
weak. The ground-state energy $E[n({\bf p})]$ of the normal
Fermi liquid is a functional of occupation numbers $n({\bf p})$
\cite{lan}, which, at $T=0$, are related to the order parameter by
the simple equation
\beq n({\bf p})=v^2({\bf p}); \,\,\,
\kappa({\bf p})=v({\bf p})\sqrt{1-v^2({\bf p})}.\eeq
Minimizing the energy $E_{gs}$ in Eq. (1) with
respect to occupation numbers and taking into account Eq. (2), we
obtain the equation
\beq \varepsilon({\bf p})-\mu=\Delta({\bf p})
\frac{1-2v^2({\bf p})} {2\kappa({\bf p})},\eeq
where the single-particle energy $\varepsilon({\bf p})$ is
determined by the equation \cite{lan},
\beq\varepsilon({\bf p})=\frac{\delta
E[n({\bf p})]}{\delta n({\bf p})},\eeq
$\mu$ is chemical potential. The
superconducting gap is given by the equation
\beq \Delta({\bf p})=-\int
V_{pp} ({\bf p}, {\bf p}_1) \kappa({\bf p}_1)
\frac{d{\bf p}_1}{4\pi^2}.\eeq
Let us assume that
the interaction $V_{pp}\to 0$. Then, the gap $\Delta({\bf p})\equiv
0$, and Eq. (3) is reduced to the equation proposed in
\cite{ks} \beq \varepsilon({\bf
p})-\mu=0,\:  {\mathrm {if}}\,\,\, 0<n({\bf p})<1;\: p_i\leq p\leq
p_f. \eeq
This equation defines a Fermi liquid of a new type for which the
order parameter $\kappa({\bf p})$ differs from zero in the
$L_{FC}$  range
of momenta $p_i\leq p\leq p_f$; the occupation numbers
$n({\bf p})=1$ and 0 outside the $L_{FC}$ range, as must be in the
normal Fermi liquid. It follows from Eq. (6) that the effective
mass $M^*_{FC}$ of quasiparticles in the fermion condensate is
infinitely large in the $L_{FC}$ range:  \beq \frac{1}{M^*_{FC}}
=\frac{1}{p}\frac{d\varepsilon(p)}{dp}=0.\eeq
The effective mass of normal
quasiparticles $M^*_{L}$ with momenta $p<p_i$ is finite and is
defined by the known equations \cite{lan}
\beq \frac{1}{M^*_{L}}
=\frac{1}{p}\frac{d\varepsilon(p)}{dp}|_{p\leq p_i}.\eeq
It follows from Eqs. (7) and
(8) that a fermion system with the fermion condensate is broken into
two quasiparticle subsystems: the dispersionless part of the single-
particle spectrum is occupied by the fermion condensate
in the momentum range $L_{FC}$  and is adjoined by the subsystem
that is occupied by quasiparticles of
finite mass with momenta $p<p_i$.
We will assume for simplicity that the fermion condensate occupies
a small part of the Fermi sphere $p_f-p_i\ll p_F$, where the Fermi
momentum is related by the common equation $p_F=(3\pi^2\rho)^{1/3}$
to the particle density $\rho$. The fermion condensate appears in an
electron system at a low density, when the effective
electron--electron interaction constant is sufficiently
large. In a common electron liquid, this constant is
directly proportional to the dimensionless parameter
$r_s=9\pi/4p_Fa_B$, where $a_B$ is the Bohr radius. For simplicity,
we will assume that it equals $r_s$. It was shown in \cite{ksz} that the
appearance of the fermion condensate occurs in a system at a certain
$r_s=r_{FC}<r_{cdw}$
and precedes the appearance of a charge-density wave,
which takes place in a two-dimensional electron liquid at
$r_{cdw}\simeq 6-8$, \cite{sns}. Thus, the
FCQPT occurs at $T=0$ when the parameter
$r_s$ attains its critical value $r_{FC}$ and represents a quantum phase
transition. At $r_s>r_{FC}$ and $r_s-r_{FC}\ll r_{FC}$, the region
$p_f-p_i$ occupied by the fermion condensate is
$(p_f-p_i)/p_F\sim r_s-r_{FC}$.
This estimate is confirmed by calculations for simple
models \cite{ksk,dkss}.

Because the order parameter of the FCQPT
is $\kappa({\bf p})$, the maximum value of the
superconducting gap $\Delta_1$ in a system with the fermion condensate
$\Delta_1\sim V_{pp}$,
as it follows from Eq. (5). It is pertinent to note that
$\kappa({\bf p})$ is determined in this case by the relatively strong
particle--hole interaction or by the Landau amplitudes $F_L$.
Therefore, the perturbation of the parameter
$\kappa({\bf p})$ can be neglected in
the first order in $V_{pp}/F_L\ll 1$. It is self-evident that we
assume the $L_{FC}$ range to be sufficiently large, so that its
perturbation is small compared with the size of this range.
Considering that $T_c\simeq \Delta_1/2$  in the
weak-coupling theory of superconductivity \cite{dkss,lp}, we obtain high
$T_c$  values for systems with the fermion condensate \cite{ks}. At
the same time, the single-particle spectrum in the range $L_{FC}$
occupied by the fermion condensate will be disturbed by the
interaction $V_{pp}$. This perturbation is quite notable for the
effective mass, because the value of $1/M^*_{FC}$ becomes finite.
Simultaneously, the perturbation of the single-particle spectrum at
$p<p_i$, as well as the perturbation of the effective mass
$M^*_L$, can be neglected.

Let us use Eq. (3) for
calculating $M^*_{FC}$ by differentiating both sides of this equation
with respect to momentum $p$ at $p=p_F$
\beq \frac{p_F}{M^*_{FC}}\simeq
\frac{\Delta_1}{4\kappa^3(p)}\frac{1}{p_f-p_i}
=\frac{2\Delta_1}{p_f-p_i}.\eeq
When obtaining Eq. (9),
we took into account the facts that $\kappa(p)=1/2$,
at $p=p_F$, the gap $\Delta({\bf p})$ has a maximum
at the Fermi surface, and, hence, its derivative there
equals zero. The derivative $dv(p)/dp$ was calculated
with the use of Eq. (2) and the simple estimate
$dn(p)/dp\simeq-1/(p_f-p_i)$. We may conclude that the electron
system with the fermion condensate in the superconducting state is, as
before, characterized by two effective masses, and that the
single-particle dispersion at $p\sim p_F$ can be approximated by two
straight lines. Let us estimate the binding energy $E_0$ at which
these lines intersect. Multiplying both sides of Eq. (9) by the
difference $p_f-p_i$, we obtain
\beq E_0\simeq\frac{(p_f-p_i)p_F}{M^*_{FC}}\simeq 2\Delta_1. \eeq
It follows from this equation
that the intersection point of the two straight lines approximating
the spectrum does not depend on the difference $p_f-p_i$, although
the effective mass $M^*_{FC}$ is proportional to this difference.
The calculation of $M^*_{FC}$ at $T\to T_c$ is completely
similar to the preceding; one should only take into account that now
\cite{lp},
\beq v^2({\bf p})=\frac{n({\bf p})-f({\bf p})}{1-2f({\bf p})},\eeq
where
\beq f({\bf p})=\frac{1}{1+\exp(E({\bf p})/T)};\,\,
E({\bf p})=\sqrt{(\varepsilon({\bf p})-\mu)^2
+\Delta^2({\bf p})}.\eeq
With regard to the facts that the function $f({\bf p})$
has a maximum at $p=p_F$ (and its derivative equals zero there) and
$E({\bf p})\ll T$, simple transformations of Eqs. (11) and (12) give
\beq\frac{d(v^2(p))}{dp}\simeq -\frac{2T}{E(p)(p_f-p_i)}.\eeq
Differentiating both sides of Eq. (3) with respect to momentum and
taking into account Eq. (13), we obtain
\beq \frac{p_F}{M^*_{FC}}\simeq \frac{4T}{p_f-p_i}.\eeq
It directly follows from Eq. (14) that
\beq E_0\simeq\frac{(p_f-p_i)p_F}{M^*_{FC}}\simeq 4T. \eeq
Considering that $2\Delta_1\simeq T_c$, we conclude by
comparing Eqs. (10) and (14) that the effective mass $M^*_{FC}$ and
$E_0$ weakly depend on temperature at $T\leq T_c$.

It follows from the
above consideration that the form of the single-particle spectrum
$\varepsilon({\bf p})$ and the order parameter $\kappa({\bf p})$ are
determined by the FCQPT and,
therefore, their forms are universal.  Actually, the amplitudes $F_L$
define only the region $L_{FC}$ occupied by the condensate after
the point of the FCQPT.
These amplitudes are determined by the properties of
the system under consideration, which already include
the contribution from impurities, phonons, and other
collective excitations. Finally, we may conclude that a
system with the fermion condensate is characterized by a
universal form of the single-particle spectrum and possesses quantum
protectorate features at $T\leq T_c$.

We now turn to the description of the system at
$T>T_c$, which is given by the equation of the Fermi-liquid
theory \cite{lan},
\beq \frac{\delta (F-\mu N)}{\delta
n({\bf p},T)}=\varepsilon({\bf p},T)-\mu(T)-
T\ln\frac{1-n({\bf p},T)}{n({\bf p},T)}=0.\eeq
Here, $F$ is free energy, which, as well as energy $E$, is a
functional of occupation numbers $n({\bf p},T)$.
The occupation numbers now depend on momentum and temperature, and
the quasiparticle energy $\varepsilon({\bf p},T)$ is defined by Eq.
(4). Assuming that $T_c=0$ and $T\to 0$ in Eq. (16) and that the
occupation numbers differ from zero and unity in the range $L_{FC}$,
we obtain that the term $T\ln(...)\to 0$, and Eq. (16) is reduced to
Eq. (6) for the fermion condensate \cite{ks}. If the interaction
$V_{pp}=0$, the FCQPT is absent at any
finite temperature. Actually, as shown above, the order parameter
$\kappa({\bf p})$ after the point of the FCQPT
differs from zero in the region $L_{FC}$, and the gap
$\Delta({\bf p})\equiv 0$. From here, it is clear that the critical
temperature of this transition equals zero.  However, a trace of this
quantum phase transition persists in its radical effect on the
properties of the system up to temperatures $T\ll T_f$, where $T_f$ is
a temperature at which the effect of this phase transition
disappears. For example, the system entropy can be taken as such a
property, resulting in the estimate
\cite{dkss},
\beq \frac{T_f}{\varepsilon_F}\sim\frac{p_f^2-p_i^2}{p^2_F}
\sim\frac{\Omega_{FC}}{\Omega_F},\eeq
where $\Omega_{FC}\sim (p_f-p_i)p_F$ is the
volume occupied by the fermion condensate,
$\Omega_F$ is the volume of the
Fermi sphere, and $\varepsilon_F$ is the Fermi energy. Taking into
account that the occupation numbers at
$T\ll T_f$ are defined by Eq.
(6) and $n({\bf p},T)=n({\bf p})$, we obtain from Eq. (16)
\beq \varepsilon({\bf p},T)-\mu(T)=
T\ln\frac{1-n({\bf p})}{n({\bf p})} \simeq T\frac{1-2n({\bf
p})}{n({\bf p})}|_{p\simeq p_F}. \eeq
Differentiating both sides of Eq. (18) with respect to momentum $p$
and using the estimate $dn(p)/dp\simeq -1/(p_f-p_i)$,
we obtain the
approximate value for the effective mass
\beq \frac{p_F}{M^*_{FC}}\simeq \frac{4T}{p_f-p_i}|_{T\ll T_f}.\eeq
Multiplying both sides of Eq. (19) by the difference
$p_f-p_i$, we obtain for the parameter
\beq E_0\simeq 4T.\eeq
Equations (19) and (20) indicate that the mass $M^*_{FC}$ and
the energy $E_0$ start to depend on temperature at $T_c\leq T\ll T_f$.
However, this dependence is very weak at $T\simeq T_c$, as is evident from
a comparison of Eqs. (14), (15), (19), and (20). We may conclude that
the system under consideration still possesses quantum protectorate
features at these temperatures, because the spectrum of the system
is determined by solutions of Eq. (6) and temperature. It is
evident from Eqs. (18), (19), and (20) that this spectrum has a
universal character and is weakly affected by phonons, collective
states, etc.

We now turn to the description of the experimental data
\cite{vall,blk,krc} using the results presented above. We return to
the consideration of an electron system on a square lattice.
Experimental studies showed that the Fermi surface in the case of
the Bi$_2$Sr$_2$CaCu$_2$O$_{8+\delta}$ metal has a shape of an
approximately regular circle with the center at the point
$(\pi,\pi)$ of the Brillouin zone filled with hole states \cite{mkf}.
A Van Hove singularity is located in the vicinity of the $(\pi,0)$
point, and an almost dispersionless section of the spectrum is
observed in this region (see, for example, \cite{mkf}). This allows the
suggestion to be made that the fermion condensate is disposed in the
vicinity of this point \cite{kcs}.
The straight line $Y\Gamma$, which is known as
the line of zeros of the Brillouin zone, passes through the points
$(\pi,\pi)$ --- $(0,0)$ at an angle of $\pi/4$ to the straight line
$Y \bar{M}$
passing through the points $(\pi,\pi)$ --- $(\pi,0)$. The density of
states attains a minimum at the point of intersection of the
$Y\Gamma$ line and the Fermi surface. The single-particle spectrum
was measured along the lines parallel to $Y\Gamma$ \cite{blk} and
$Y\bar{M}$ \cite{krc},
from the line of zeros to the $Y\bar{M}$ line.  As a result, it was shown
that the parameter $E_0$ is constant for a given sample; that is, it
does not depend on the angle
$\phi$, reckoned from the line of zeros to $Y\bar{M}$.
The angle (kink) between the straight line characterizing the part of
the spectrum with the binding energy lower than $E_0$ and the
straight line related to the spectrum with the binding energy higher
than $E_0$ grows with increasing $\phi$ and with decreasing doping level
\cite{blk,krc}. This general pattern is retained at $T>T_c$ \cite{blk}.

To describe these experimental data, we assume the following model:
the volume of the fermion condensate $\Omega_{FC}$ depends on the
angle $\phi$, $\Omega_{FC}(\phi)\sim (p_f(\phi)-p_i(\phi))p_F$,
increases with increasing $\phi$, and attains a maximum at $(\pi,0)$.
Along with that, $r_s$ grows with decreasing doping level and exceeds
the critical value $r_{FC}$ in the optimal doping region.  Note that
the values of $r_s$ corresponding to the optimal doping level are
close to $r_{FC}$ \cite{ksz,ars,ams}, whereas strong fluctuations of
the charge density or charge-density waves are observed in undoped
samples \cite{grun}. From here, we may conclude that the formation of
the fermion condensate in copper oxides is a quite determinate process
stemming from the general properties of a low-density electron
liquid.

It follows from
Eq. (15) that the energy $E_0$ does not depend
on the angle $\phi$ at $T\leq T_c$. It also follows from Eq. (14)
that the kink increases with increasing $\phi$, because the effective mass
linearly depends on the difference $(p_f(\phi)-p_i(\phi))$. Comparing
Eqs. (9), (10), (14), and (15), one can conclude that these
properties weakly depend on temperature at $T\leq T_c$. Equations (19)
and (20) demonstrate that this behavior persists at $T_c\leq T$;
however, a temperature dependence appears. According to experimental
data, $E_0\simeq (50 - 70)$ meV \cite{blk,krc}, which is in
agreement with Eqs. (10) and (20), because $E_0\simeq 2\Delta_1$ in
these materials. The volume of the phase condensate $\Omega_{FC}$ in
our model grows with increasing $r_s$, and the mass
$M^*_{FC}$ correspondingly increases, as evident from Eqs. (9) and
(14).  Hence, the dispersion kink in the single-particle spectrum
must increase with decreasing doping level. Because $E_0\simeq
2\Delta_1$, the kink point must shift towards higher binding energies
as the doping level decreases.  All these results are in a good
agreement with experimental data \cite{blk,krc}.

The lineshape of a single-particle excitation is
another important characteristic property of this excitation that
can be measured experimentally. The lineshape $L(q,\omega)$ is a
function of two variables. Measurements carried out at a fixed
binding energy $\omega=\omega_0$, where $\omega_0$ is the energy of
the single-particle excitation under study, determine the lineshape
$L(q,\omega=\omega_0)$, as a function of momentum $q$ \cite{vall}.
As shown above, the effective mass $M^*_{FC}$ is finite
if $T>0$, or $\Delta\neq 0$. Therefore, the system behaves as a
normal liquid characterized by a certain effective mass at energies
$\omega<4T$ (or $\omega<2\Delta_1$, if
$T<T_c$). Quasiparticles with energies of the order of temperature
will be involved in rescattering processes, which determine the width
of the single-particle excitation. As follows from Eq. (20), these
are precisely the quasiparticles with mass $M^*_{FC}$, which leads
to the width of the order of $T$ \cite{dkss}. It was this behavior that was
observed in experiments on measuring the lineshape at a fixed
energy, when well-defined quasiparticles at the Fermi level were
found even in the region of the $(\pi,0)$ point \cite{vall}. The
lineshape can be determined differently as a function of energy
$\omega$ at a fixed momentum $q$ \cite{nor}. At small $\omega$, the
line will have a characteristic maximum and width as well as in the
case of fixed energy $\omega$ . At energies
$\omega\geq 4T$ (or $\omega\geq 2\Delta_1$ if $T<T_c$),
quasiparticles of mass $M^*_{L}$ will come into play, which will lead
to a growth of the function that determines the lineshape.  Thus,
this line will have a characteristic shape: a maximum, then a
minimum, and then again a flat maximum.  This result is in a
qualitative agreement with the experiment \cite{ars,ams}.
On the other hand, one may follow the procedure suggested
in \cite{krc}, using the Kramers-Kr\"{o}nig transformation to
construct the imaginary part of the self-energy starting with the
real one. As we have seen above, the real part of the self-energy
in systems with FC can be represented by two straight lines
intersecting near the point $E_0$ and
characterized by the two effective masses.
As a result, the lineshape $L(q=q_0,\omega)$ of the
quasiparticle peak as a function of the binding energy $\omega$
possesses a complex peak-dip-hump structure \cite{krc} directly
defined by the existence of the two effective masses $M^*_{FC}$
and $M^*_L$ \cite{ars}.

V.R.S. is grateful to the Racah Institute of Physics, Hebrew
University of Jerusalem for hospitality.  This work was supported in
part by the Russian Foundation for Basic Research, project no.
01-02-17189.

\end{multicols}

\end{document}